\documentclass[11pt,twoside]{book} 
\usepackage{asp2008s}
\usepackage{times}
\usepackage{lscape}
\usepackage{epsf}
\usepackage{index}

\usepackage{graphicx}
\usepackage{amssymb}
\usepackage{longtable}
\usepackage[figuresright]{rotating}
\usepackage{multirow}

\newcommand{\simless}{\mathbin{\lower 3pt\hbox {$\rlap{\raise 5pt\hbox{$\char'074$}}\mathchar"7218$}}}

\mainmatter

\begin{document}

\title{The W40 Cloud Complex}   

\author{Steven A. Rodney}
\affil{Institute for Astronomy, University of Hawaii\\
	2680 Woodlawn Dr., Honolulu, HI 96822, USA}

\author{Bo Reipurth}
\affil{Institute for Astronomy, University of Hawaii\\
640 N. Aohoku Place, Hilo, HI 96720, USA}    

\begin{abstract} 
The W40 complex is a nearby site of recent massive star formation
composed of a dense molecular cloud adjacent to an HII region that
contains an embedded OB star cluster.  The HII region is beginning to
blister out and break free from its envelope of molecular gas, but our
line of sight to the central stars is largely obscured by intervening
dust.  Several bright OB stars in W40 - visible at optical, infrared,
or cm wavelengths - are providing the ionizing flux that heats the HII
region. The known stellar component of W40 is dominated by a small
number of partly or fully embedded OB stars which have been studied at
various wavelengths, but the lower mass stellar population remains
largely unexamined. Despite its modest optical appearance, at 600~pc
W40 is  one of the nearest massive star forming regions, and with a UV
flux of about 1/10th of the Orion Nebula Cluster, this neglected
region deserves detailed investigation.
\end{abstract}


\section{Overview}

The star forming region known as W40 consists of three interrelated
components. First, there is the cold molecular cloud, which is
designated using Galactic coordinates as G28.8+3.5, following
\citet{gos1970}.  This dark cloud has an angular extent on the order
of one degree, and is centered around a dense molecular core with a
diameter of approximately $20\arcmin$, identified as TGU~279-P7 in the
recent extinction atlas of \citet{dob2005}.

Adjacent to this molecular cloud is the large blister HII region
denoted as W40 \citep{wes1958}, and also labeled as S64 by
\citet{sha1959}, as RCW~174 by Rodgers, Campbell, \& Whiteoak (1960),
or LBN~90 in the Lynds (\citeyear{lyn1965}) catalog of bright nebulae.
The W40 HII region is centered on J2000 coordinates $18^h31^m29^s$,
$-2\deg05\farcm4$, and has a diameter of $\sim6\arcmin$.  The space
between the hot, ionized HII region and the cold molecular cloud is
marked by a thin CII region and an accompanying neutral interface
\citep{val1987}.

\begin{figure}[tbp]
\plotfiddle{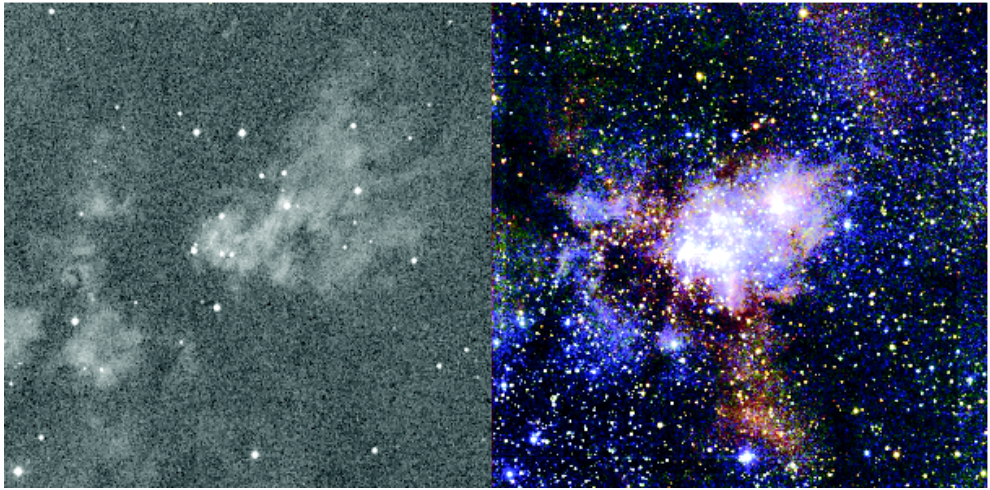}{6.3cm}{0.0}{85.0}{85.0}{-185.0}{0.0}
\caption{The W40 region seen on (left) the red Digitized Sky Survey
  and (right) at JHK with 2MASS. The field is approximately
  17$\times$17 arcmin, corresponding to a little less than 3 pc on the
  side at 600~pc. North is up and east is left. Courtesy K.~Getman.
}
\label{fig:dss_2mass}
\end{figure}

Lastly, the W40 region hosts an embedded stellar cluster that is
dominated by three bright IR sources \citep{zei1978}.  These bright OB
stars are the primary excitation sources for the W40 HII region
\citep{smi1985} and show evidence for substantial circumstellar
envelopes \citep{val1994}.  The W40 cluster is heavily obscured along
our line of sight by the surrounding molecular cloud, which provides
$A_V \sim10$ magnitudes of visual extinction throughout, and as much
as $A_V = 17$ mag at the dense center \citep{rey2002}. Figure
\ref{fig:dss_2mass}  shows the W40 region in the optical from the
Digitized Sky Survey and in the infrared from the 2MASS survey.

\section{Distance}

The distance to W40 has not yet been determined to any satisfactory
precision.  Measurements of the H109$\alpha$ atomic recombination line
at 0.7 km~s$^{-1}$ \citep{rei1970}, the $\lambda$18 cm OH absorption
line at 6.3 km~s$^{-1}$ \citep{dow1970} and the 21 cm HI absorption
line along the line of sight at 7.2 km~s$^{-1}$ \citep{rad1972}
collectively suggest a rough kinematic distance estimate of 300-900 pc
based on the assumption that the cloud is in circular motion about the
Galactic center.  Estimates based on the radio/IR continuum of W40's
stellar component allow for a similar range from 400 pc
\citep{cru1982} to 700 pc \citep{smi1985}.  Using OH line measurements
with a unique distance determination technique, \citet{kol1983}
calculated a distance of 600 pc.

Adopting a conservative mean of 600 pc places W40 at a distance of 37
pc above the Galactic plane.  The dense central region of the
molecular cloud is then $\sim3.5$ pc in diameter, and the width of the
HII region measures $\sim1$ pc.

\begin{table}[!p]
\caption{Radio Observations}
\smallskip
\begin{center}
{\small
\begin{tabular*}{\textwidth}{@{\extracolsep{-2mm}}ccc@{\extracolsep{3mm}}l@{\extracolsep{1.5mm}}cl}
\tableline
\noalign{\smallskip}
	      Freq.
	    & Wave.
	    & Flux$^a$
	    & Telescope
	    & HPBW
	    & Reference \\

	     (GHz)
	    &(cm)
	    &(Jy)
	    & Location$^b$
	    &(arc min)
	    & \\ 
\noalign{\smallskip}
\tableline
\tableline
\noalign{\smallskip}
\multicolumn{6}{c}{Continuum Observations} \\
\noalign{\smallskip}
\tableline
\noalign{\smallskip}

0.328 & 91.5 & 34 &  VLA-C - 3.4 km & 46$\arcsec$ & \cite{val1991} \\
0.408 & 73.5 & 34 &  Molonglo - 1.6 km & $1.4\times 2.5$ & \cite{kes1968} \\
0.408 & 73.5 & 25 &  Molonglo - 1.6 km & $2.6\times 3.4$ & \cite{sha1970a} \\
0.610 & 49.2 & 35 &  Jodrell Bank - 76 m & 30 & \cite{mor1965}  \\
0.960 & 31.3 & 25 &  Owens Valley - 27 m & 48 & \cite{wil1963}  \\
1.390 & 21.6 & 50 &  Dwingeloo - 25 m & 34.2 & \cite{wes1958}  \\
1.400 & 21.4 & 33 &  Green Bank - 91 m & 10  & \cite{fel1972} \\
1.414 & 21.2 & 36 &  Green Bank - 91 m & 10 & \cite{alt1970} \\
1.465 & 20.5 & 30 &  VLA-C - 3.4 km & 12$\arcsec$ & \cite{val1991} \\
2.695 & 11.1 & 34 & Green Bank - 43 m & 11 & \cite{alt1970} \\
4.86  & 6.17 & - & VLA-B - 10.6 km & $\sim2\arcsec$ &\cite{mol1998} \\
5.000 & 6.00 & 35 & Fort Davis - 26 m & 11 & \cite{alt1970} \\
5.000 & 6.00 & 32 & Parkes - 64 m & 4.0 & \cite{gos1970} \\
5.009 & 5.99 & 35 & Green Bank - 43 m & 6.5 & \cite{rei1970} \\
14.94 & 2.0 & - & VLA-B - 10.6  km & $\sim2\arcsec$ & \cite{mol1998} \\

\noalign{\smallskip}
\tableline
\end{tabular*}

\begin{tabular*}{\textwidth}{@{\extracolsep{-1mm}}ccc @{\extracolsep{3mm}}l@{\extracolsep{1.8mm}} cl}
\noalign{\smallskip}
\multicolumn{6}{c}{C100$\alpha$ / C125$\alpha$  Observations} \\
 (GHz) &(cm) &(mJy) & &(arc min) \\

\noalign{\smallskip}
\tableline
\noalign{\smallskip}

1.425 & 21.0 & 84 $^c$ & Effelsberg - 100 m & 8.5 & \cite{pan1977} \\
3.327 & 9.01 & 156 & Algonquin - 46 m & 8.2 & \cite{val1987} \\
6.482 & 4.62 & 86  & Algonquin - 46 m & 4.3 & \cite{val1987} \\

\tableline
\noalign{\smallskip}

\multicolumn{6}{l}{$^a$ Value quoted is the integrated flux density of the HII region continuum} \\
\multicolumn{6}{l}{$^b$ Single dish diameter in meters; longest interferometer baseline in km} \\
\multicolumn{6}{l}{$^c$ Deduced from their values of brightness temperature and angular size \citep{val1987}} \\
\multicolumn{6}{l}{{\sc Note} $-$ Adapted from Table 1 of \cite{val1987} } \\

\end{tabular*}
}
\label{tab:radio_obs}

\caption{Millimeter Observations}
\smallskip
{\small
\begin{tabular}{ @{\extracolsep{-0.2mm}}lcl@{\extracolsep{-2mm}}cl}
\tableline
\noalign{\smallskip}

	       Molecular
	    &  Wave
	    &  Telescope
	    &  HPBW
	    &  Reference \\

	      Transition
	    & (mm)
	    & Location
	    & (arc sec)
	    &  \\

\noalign{\smallskip}
\tableline
\tableline
\noalign{\smallskip}

HCO{\tiny$^+$} J=1-0   & 3.36 & Crimean Astr.Obs. - 22 m & 40 & \cite{pir1995} \\
C$^{18}$O J=1-0 & 2.73 & Pico Veleta - 30 m & 22 & \cite{val1992} \\
$^{13}$CO J=1-0 & 2.72 & Pico Veleta - 30 m & 22 & \cite{val1992} \\
$^{12}$CO J=1-0 & 2.6 &  Kitt Peak - 11 m & 1.2 & \cite{wil1974} \\
$^{12}$CO J=1-0 & 2.6 & McDonald Obs. - 5 m & 2.3 & \cite{zei1978} \\
$^{12}$CO J=1-0 & 2.6 & Holmdel - 7 m & 1.7 & \cite{bli1982} \\
$^{12}$CO J=1-0 & 2.6  & Purple Mountain - 13.7 m & 50 & \cite{zhu2006} \\
$^{13}$CO J=2-1 & 1.36 & Pico Veleta - 30 m & 12 & \cite{val1992} \\
$^{13}$CO J=2-1 & 1.36 & KOSMA - 3 m & 120 & \cite{zhu2006} \\
C$^{18}$O J=2-1 & 1.37 & Kitt Peak - 12 m & 32 & \cite{val1992} \\
$^{13}$CO J=2-1 & 1.36 & Kitt Peak - 12 m & 32 & \cite{val1992} \\
$^{13}$CO J=3-2 & 0.9  & KOSMA - 3 m & 80 & \cite{zhu2006} \\
$^{12}$CO J=3-2 & 0.87 & KOSMA - 3 m & 80 & \cite{zhu2006} \\

\tableline
\end{tabular}
}
\end{center}
\label{tab:mm_obs}
\end{table}

\section{The HII Region and Molecular Cloud}

After the initial discoveries of the W40 radio emission nebula and its
associated HII region around 1960, subsequent radio detections were
recorded over a wide range of frequencies.  By the early 1970s the
radio spectrum was well-constrained from 408 MHz to 5 GHz.  Table
\ref{tab:radio_obs} summarizes the radio observations of the HII
region's free-free continuum and carbon recombination lines in the CII
shell.
The median flux
density of the radio continuum in the HII region is 34 Jy. The most
detailed radio continuum analysis to date is from the VLA observations
of \citet{val1991}, which show a linear increase in the flux density
across the HII region from the surrounding interstellar medium in the
NE to the molecular cloud edge in the SW. Their maps also suggest the
presence of a $1.7\arcmin$ shell around the star IRS2a.

The TGU~279 molecular cloud has also been extensively mapped at
millimeter wavelengths.  These observations are summarized in Table
\ref{tab:mm_obs}.  The properties of the molecular cloud derived from
these observations are strongly model-dependent. \citet{val1992} find
that the total mass of the central cloud core out to a radius of
0.4~pc is $\sim100M_{\sun}$, and the density profile goes as $r^{-1}$.
Zhu, Wu, \& Wei (2006)
measured the mass of the cloud core at the center of the outflow to be
188 or 319 M$_{\sun}$, depending on the transition used for the
estimate.
The larger cloud complex probably has a mass of order $10^4 M_{\sun}$.
Additionally, the magnetic field strength within W40 has been
determined from Zeeman measurements of the 1665 and 1667 MHz OH lines
to be B=$-$14.0 $\pm$ 2.6 $\mu$G \citep{cru1987}.

\begin{figure*}[tbp]
  \centering
  \includegraphics[width=0.9\textwidth,draft=false]{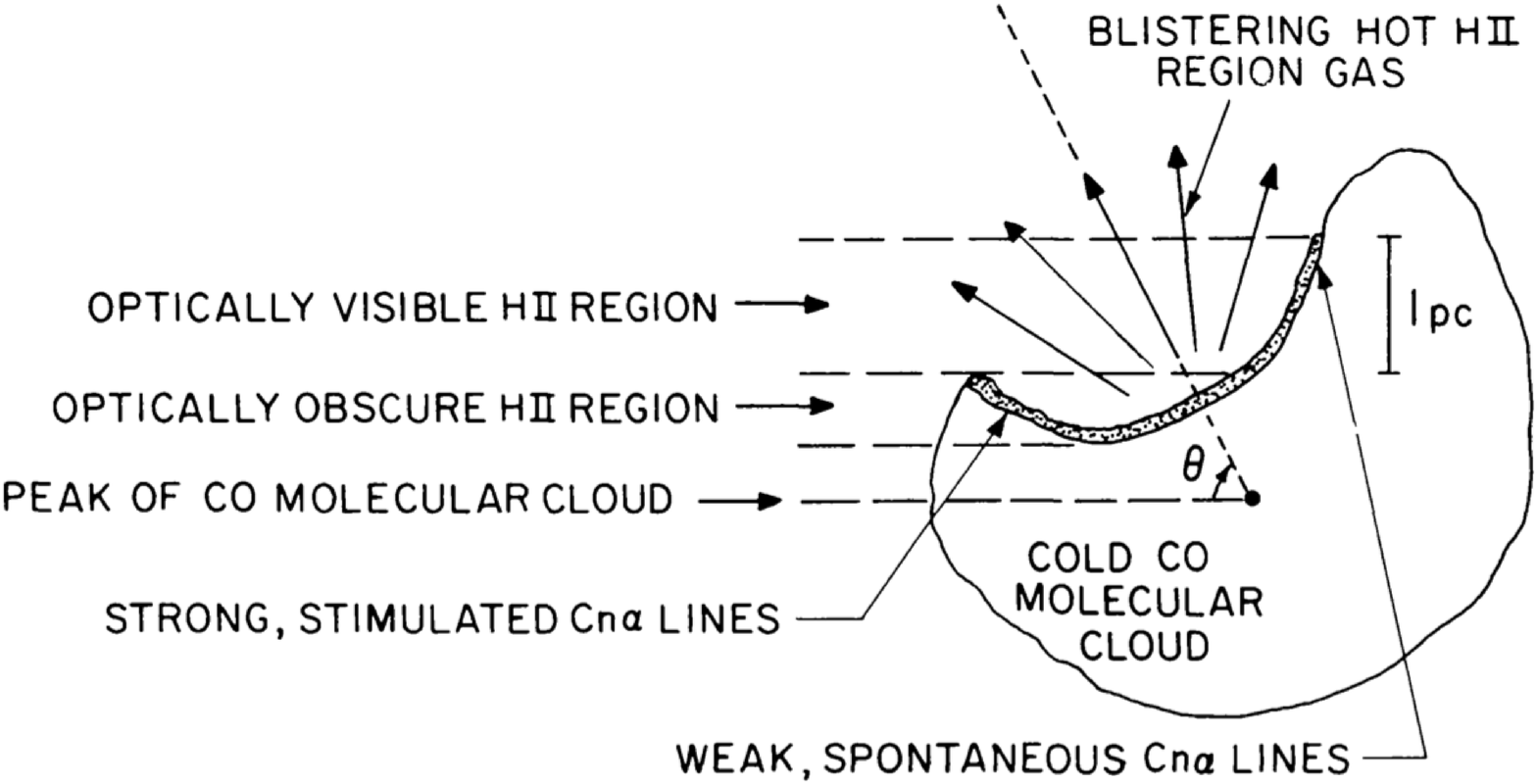}
  \caption{ Sketch of the relative positioning of the warm CII region
  ($\sim10^2$ K) between the hot blister HII region ($\sim10^4$ K) to the
  north and the cold CO molecular cloud ($\sim10$ K) to the south.  Taken
  from \citet{val1987}}.
  \label{fig:cartoon}
\end{figure*}

The first detailed infrared examination of the W40 region was carried out by
\citet{zei1978}. These authors find that the carbon recombination lines
and CO molecular lines have velocities that are offset from the published
hydrogen recombination lines by $\sim$4 km~s$^{-1}$. In addition, the bright
IR sources are spatially displaced from the CO peak, suggestive of a blister
HII region expanding away from the dense molecular cloud core.

In 1982, \citeauthor{cru1982} built on this conclusion by reviewing the
available radio recombination line
measurements, supplemented with their own molecular and H$\alpha$ maps.
These authors note the presence of two molecular components
with velocities that differ by about 3 km~s$^{-1}$, and suggest that they
correspond to the compressed gas on either side of the HII region, along our
line of sight.

\citet{val1987} took measurements of Cn$\alpha$ recombination lines in the
warm neutral interface
between the hot HII region and the cold molecular cloud. Fitting a model to
their observations, they were able to constrain the properties of the thin CII
shell and surmise that the stimulated Cn$\alpha$ line emission must be
driven by a high background temperature provided by the partially obscured
HII region.

With these observations we may begin to draw a coherent picture of the
history and current state of the W40 molecular cloud complex.  Star formation
in this region was initiated perhaps several million years ago by an
external shock which caused a compression of the molecular gas and a distinct
shift in the velocity of some of the molecular hydrogen. The
subsequent formation of a number of O and B stars created the HII region,
with a center offset by $\sim2 \arcmin$ to the northeast of the molecular
cloud core. This HII region began to expand and ionize the surrounding gas,
eventually breaking through to reveal some patchy H$\alpha$ emission.
A screen of molecular hydrogen and dust still remains along our line
of sight, obscuring most of the embedded  HII region, and the thin
shell between these cold and hot media is
dominated by stimulated Cn$\alpha$ emission.  A schematic
representation of this geometry, taken from \citet{val1987}, is shown
in Figure \ref{fig:cartoon}.

\section{The Stellar Sources}

\begin{table}[tb]
\caption{Stellar Sources in the W40 Cluster}
\smallskip
\begin{center}
{\small
\begin{tabular}{llllll}
\tableline
\noalign{\smallskip}

	    \multicolumn{3}{c}{ID: W40-\ }
	    & R.A. (2000)
	    & Dec. (2000)
	    & References \\

\noalign{\smallskip}
\tableline
\tableline
\noalign{\smallskip}

VLA 1  & &        & 18$^h$31$^m$14$\fs$8 & -02$\deg$03$\arcmin$50$\arcsec$ & 4 \\
VLA 2  & &        & 18 31 15.3 & --02 04 15 & 4 \\
& IRS 2b & OS 2b  & 18 31 22.3 & --02 05 32 & 2 \\
VLA 3  & &        & 18 31 22.3 & --02 06 19 & 4 \\
VLA 4  & &        & 18 31 23.2 & --02 06 18 & 4 \\
VLA 5  & &        & 18 31 23.6 & --02 05 35 & 4 \\
VLA 6  & &        & 18 31 23.6 & --02 05 28 & 4 \\
  & IRS 3a & OS 3a & 18 31 24.0 & --02 04 11 & 1, 2, 3 \\
VLA 7 & IRS 2a & OS 2a  & 18 31 24.0 & --02 05 30 & 1, 2, 3, 4 \\
&  &  OS 2c		& 18 31 25.2 & --02 05 10 & 2 \\
 VLA 8 & IRS 1c & & 18 31 26.0 & --02 05 17 & 2, 4 \\
& & OS 4a		& 18 31 26.5 & --02 04 31 & 2 \\
VLA 9  & & & 18 31 27.3 & --02 05 04 & 4 \\
VLA 10 & & & 18 31 27.5 & --02 05 12 & 4 \\
VLA 11 & & & 18 31 27.6 & --02 05 18 & 4 \\
VLA 12 & & & 18 31 27.6 & --02 05 13 & 4 \\
VLA 13 & IRS 1d & OS 1d  & 18 31 27.7 & --02 05 10 & 2, 4 \\
VLA 14 & & & 18 31 27.7  & --02 05 20 & 4 \\
VLA 15 & IRS 1a & OS 1a & 18 31 27.8 & --02 05 22 & 1, 2, 3, 4 \\
VLA 16 & & & 18 31 28.0 & --02 05 18 & 4 \\
VLA 17 & & & 18 31 28.6 & --02 05 29 & 4 \\
VLA 18 & IRS 1b &  & 18 31 28.7 & --02 05 30 & 2, 4 \\
VLA 19 & & & 18 31 28.7 & --02 05 22 & 4 \\
VLA 20 & & & 18 31 30.2 & --02 07 18 & 4 \\

\tableline
\noalign{\smallskip}
\multicolumn{6}{l}{{\sc References} $-$ (1) \cite{zei1978}, (2) \cite{smi1985},}\\
\multicolumn{6}{l}{(3) \cite{val1991}, (4) Rodr{\'{\i}}guez, Rodney, \& Reipurth (2008, in preparation)} \\

\end{tabular}
}
\end{center}
\label{tab:stars}
\end{table}

\begin{figure}[tbp]
\centering
\includegraphics*[height=0.4\textheight]{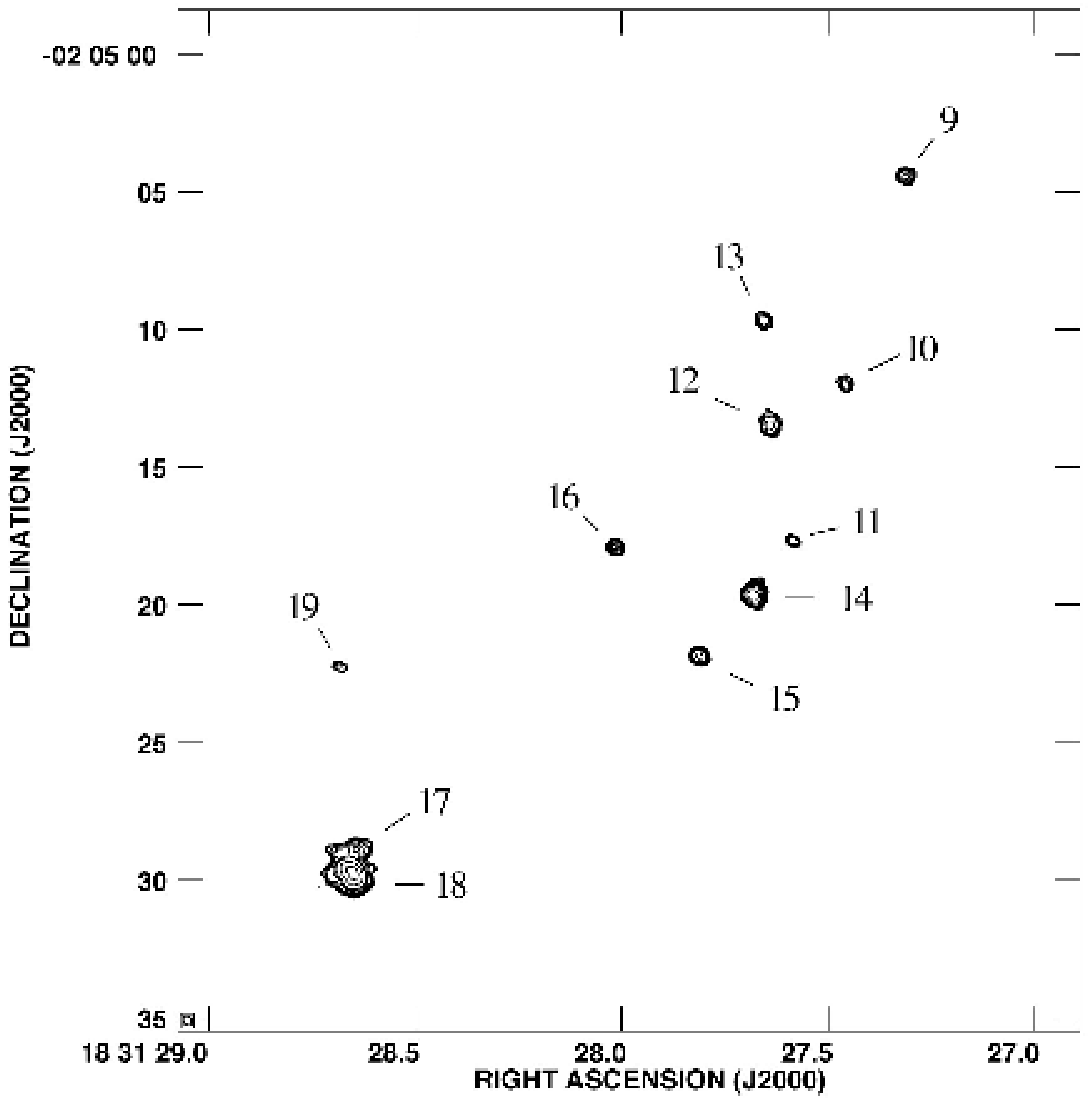}
  \caption{ VLA contour map showing the central cluster of compact radio
  sources in the W40 region. The labels correspond to the ID numbers
  indicated in Table \ref{tab:stars}.
  The majority of these sources show variability on timescales of $<$ 20
  days. From Rodr{\'{\i}}guez, Rodney, \& Reipurth (2008, in preparation).}
  \label{fig:vlamap}

\includegraphics*[height=0.4\textheight]{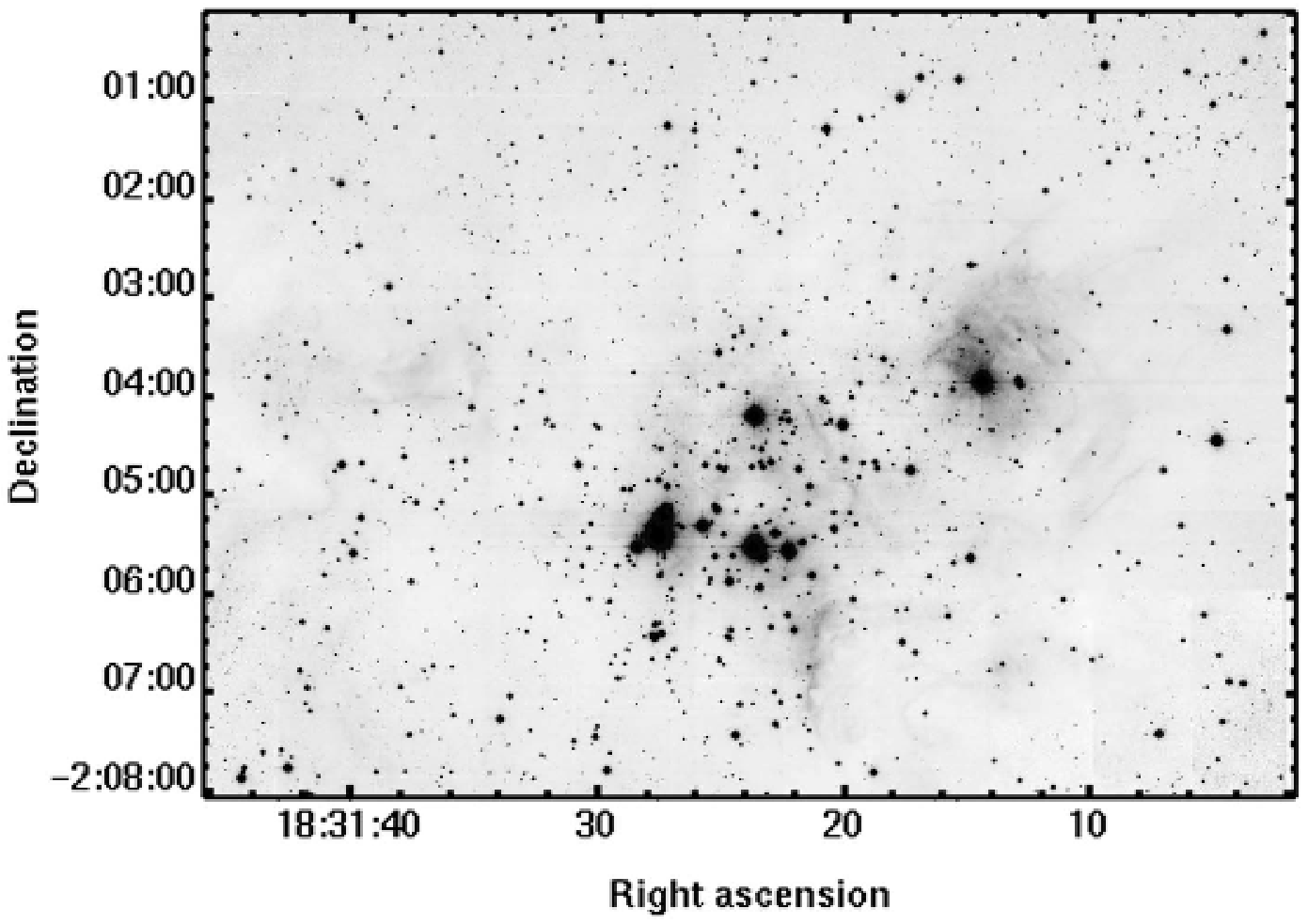}
 \caption{Composite JHK' image showing the W40 region at $\sim$2~$\mu$m. The three brightest infrared sources
  have been identified as OB stars powering the HII
  region. Coordinates are J2000.
  From Rodr{\'{\i}}guez, Rodney, \& Reipurth (2008, in preparation).}
  \label{fig:irmap}
\end{figure}

In contrast to the relatively detailed investigations of the cloud
complex, very little work has been done on the underlying stellar
population.  Three dominant IR sources in W40 were noted by
\citet{zei1978}, and subsequent observations revealed that six of the
seven brightest IR sources can be matched with optical counterparts
\citep{smi1985}.  Recent high-resolution radio continuum observations
using the VLA telescope have produced detailed maps of the W40 cluster
at 3.6~cm (Rodr{\'{\i}}guez, Rodney, \& Reipurth 2008, in
preparation).
These measurements reveal that the W40
cluster harbors 20 compact radio sources, 15 of which have IR
counterparts, and 10 of which show substantial variability in their
radio flux.  Figure \ref{fig:vlamap} shows a $\sim30\arcsec$ band of
11 of these VLA sources.  The brightest of W40's optical, IR and radio
sources are catalogued in Table \ref{tab:stars}.  The entire W40
cluster is shown in Figure \ref{fig:irmap}, which presents a composite
JHK' image of the W40 cluster taken with the University of Hawaii 2.2m
telescope.

%
%

%
%
%
%

%
%

\begin{figure}[tbp]
\centering
\includegraphics[height=0.39\textheight]{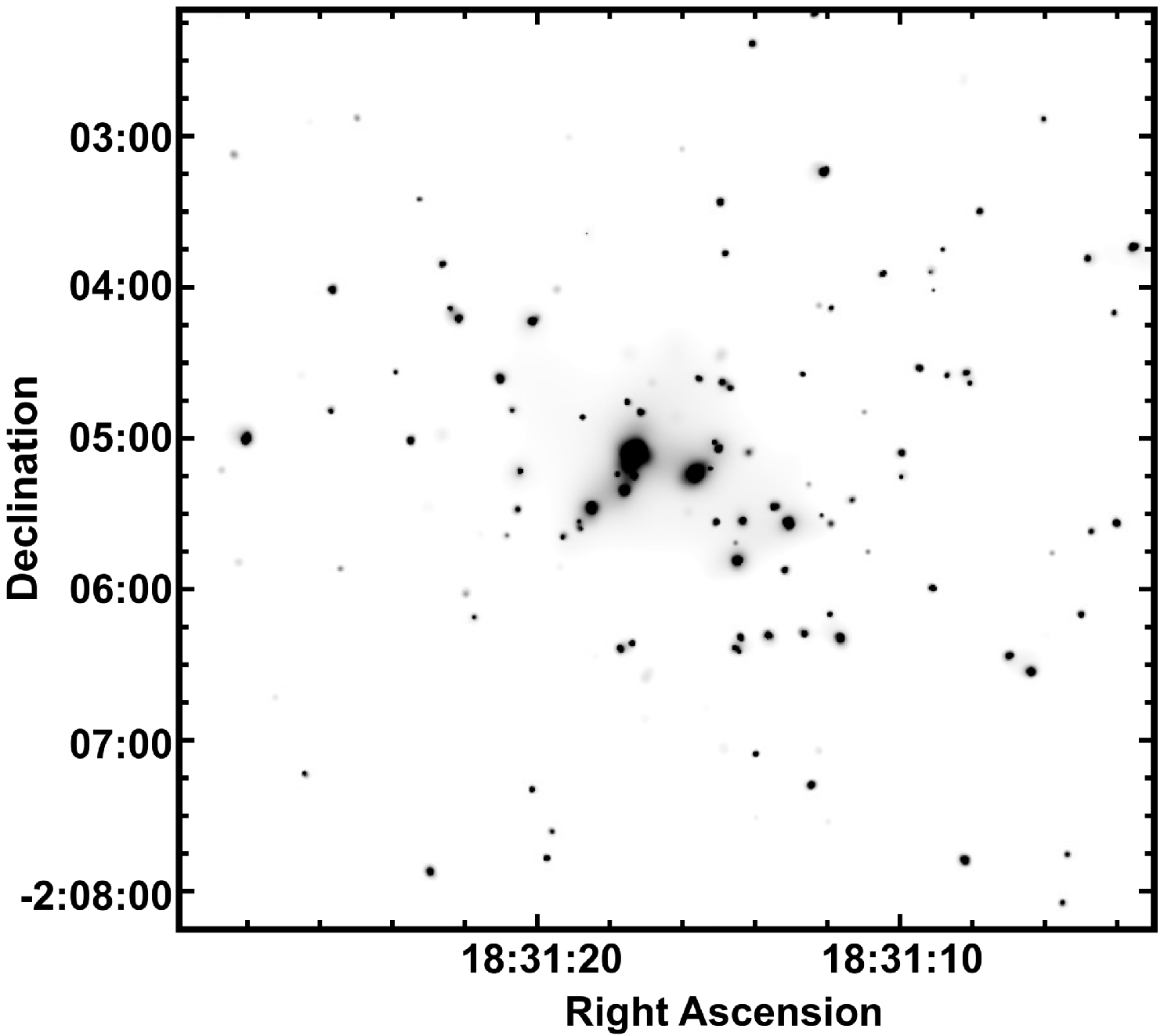}
\caption{ Close-up adaptively smoothed Chandra view of the
$6.5\arcmin \times 6\arcmin$ field around the ionizing OB stars.
Smoothing has been performed in the (0.5-8.0) keV band at the 2.5
sigma level, and gray scales are logarithmic.  Over 100 X-ray point
sources along with the diffuse X-ray emission can be seen.
Coordinates, given as J2000, are only approximate.
Courtesy K.~Getman.
}
\label{fig:chandra}
\vspace{5mm}
\includegraphics[height=0.39\textheight]{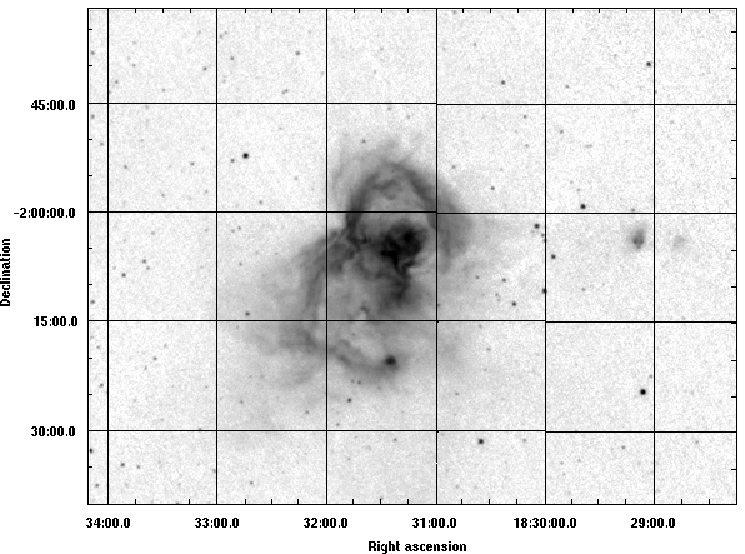}
\caption{The W40 region as seen by MSX at a wavelength of
  8~$\mu$m. The HII region is here seen with minimal impact of
  extinction, and appears as an hour-glass shaped structure about 3
  $\times$ 5 pc in extent.}
\label{fig:msx}
\end{figure}

The high energy emission of the stars in the W40 cluster is now being
examined with an X-ray study using the Chandra space telescope
(Getman, K. et~al. 2008, in preparation).
Figure \ref{fig:chandra} shows a preliminary Chandra X-ray map of the
cluster, which reveals scores of X-ray point sources coincident with
the optical and IR stellar cluster.  Some diffuse X-ray emission at
the cluster center has also been detected.

The infrared spectral energy distributions of the brightest cluster members
were examined by \citet{smi1985}, and later millimeter observations of
these sources with the James Clerk Maxwell Telescope in Hawaii have
revealed further evidence for substantial circumstellar material
surrounding IRS1a and IRS2a \citep{val1994}.  \cite{smi1985} have
shown that the circumstellar material around these sources is not an
important source of IR re-radiation in W40, but rather the diffuse
dust throughout the region is the dominant absorbing medium.
Molecular absorption lines towards IRS 1a, the brightest of the IR
sources,\footnote{Although IRS1a is the brightest IR source
in the region, Smith et al. (1985) present evidence that it is not the
most luminous.  The more heavily obscured IRS2a is considered to be
the more luminous object, and it appears to be the primary source of
ionizing radiation.}
 have been used to deduce column densities, temperatures, and
abundance ratios in the foreground region of the W40 molecular cloud
\citep{shu1999}.

The Midcourse Space Experiment (MSX) satellite has observed the W40
region, and a map obtained at 8~$\mu$m is shown in Figure
\ref{fig:msx}. Here the
full structure of the W40 region is seen with minimal interference
from obscuring dust clouds. We see that W40 consists of two
interconnected cavities, forming an hour-glass shape. The main cluster
is located just northwest of the narrow waist where the two cavities
are joined.  The total extent of the two cavities is roughly
17~$\times$~28 arcmin, which at the assumed distance of 600~pc
translates into about 3~$\times$~5~pc. Numerous point sources are
detected by MSX in W40, but only one has a good match with the
coordinates of the sources listed in Table~3: the MSX source
G028.7799+03.4978 is 2.5$\arcsec$ from IRS~2a. For this source, MSX
measured flux densities of  27.1, 41.7, 55.7 and 78.8 Jy at 8, 12, 15
and 21~$\mu$m, respectively.

\begin{figure}[tbp]
\plotfiddle{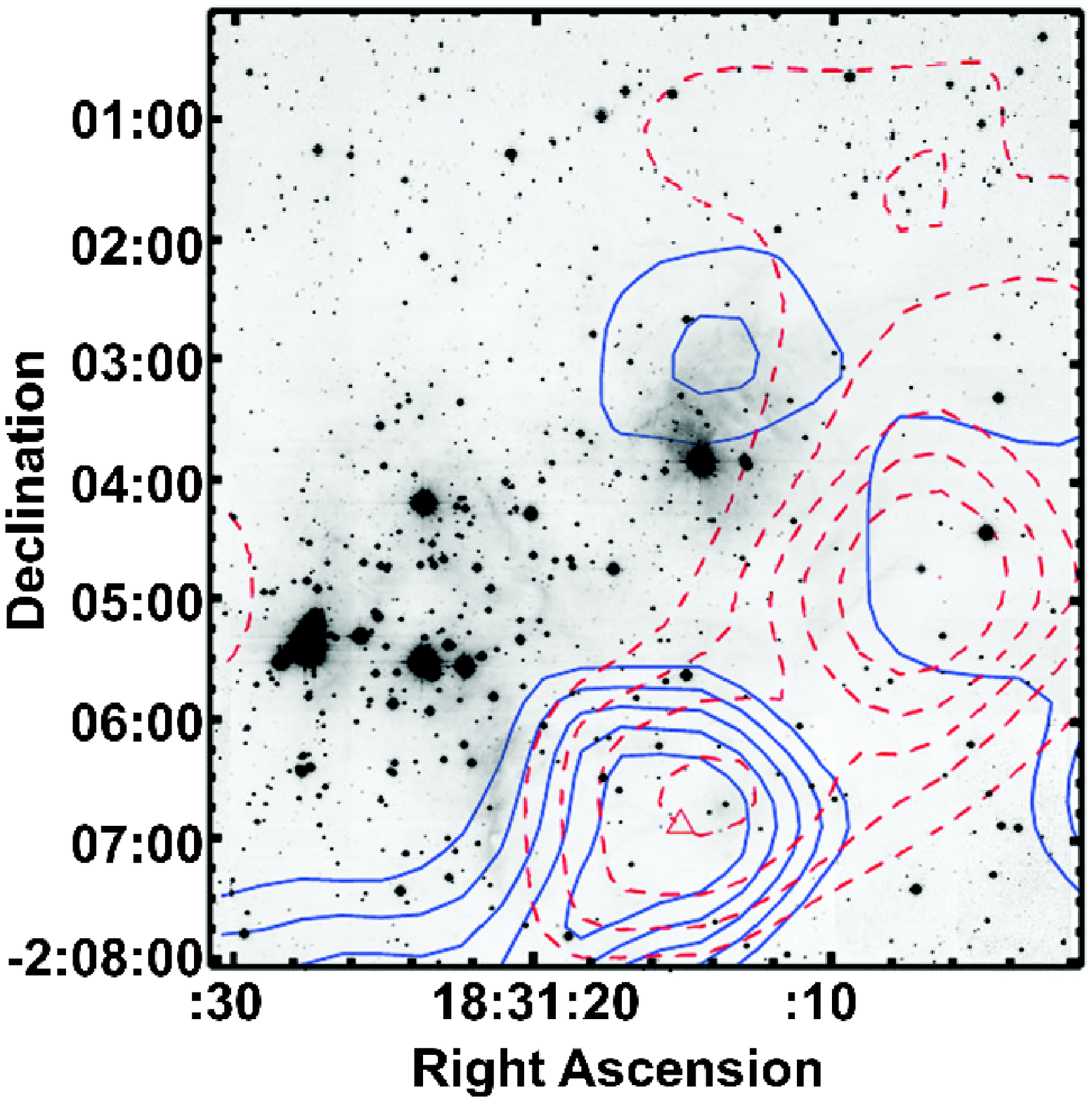}{8.4cm}{0.0}{50.0}{50.0}{-130.0}{0.0}
\caption{The molecular cloud to the southwest of the W40 cluster
  contains a weak molecular outflow discovered by Zhu et
  al. (2006). The solid line is the blue lobe, and the dotted line is
  the red lobe. A second outflow is possibly present farther to the
  north.  The millimeter observations are superposed on the 2~$\mu$m
  image from Figure~4. }
\label{fig:outflow}
\end{figure}

Zhu, Wu, \& Wei (2006) obtained $^{12}$CO J=3-2 and $^{13}$CO J=2-1 and
J=3-2 observations of the W40 region and found a molecular outflow
associated with the molecular cloud on the southwestern
rim of the HII region. The spatial position of the outflow relative to
the cluster is seen in Figure \ref{fig:outflow}.
Neither the MSX nor
the IRAS catalogues show a source at the center of the outflow, suggesting that
a very young, very embedded source is powering the outflow
activity. Clearly star formation is still taking place in the W40
region.



\vspace{0.3cm}

{\bf
Acknowledgements}. We are thankful to Konstantin Getman for
providing Figures~1 and \ref{fig:chandra} and for a thorough and very
helpful referee's report.
This work has made use of The Digitized Sky
Surveys, produced at the Space Telescope Science Institute. The images
of these surveys are based on photographic data obtained using the
Oschin Schmidt Telescope on Palomar Mountain and the UK Schmidt
Telescope. This research has made use of NASA's Astrophysics Data
System; data products from the Two Micron All Sky Survey, which is a
joint project of the University of
Massachusetts and the Infrared Processing and Analysis
Center/California Institute of Technology, funded by the National
Aeronautics and Space Administration and the National Science
Foundation; images from the MidCourse Space
Experiment; and the SIMBAD database, operated at CDS, Strasbourg,
France.
BR was supported in part by the NASA Astrobiology Institute under
Cooperative Agreement No. NNA04CC08A and by the NSF through grants
AST-0507784 and AST-0407005.


\end{document}